\documentclass[12pt,preprint]{aastex}

\newcommand{\simlt}{\lower.5ex\hbox{$\; \buildrel < \over \sim \;$}}

\slugcomment{draft date: \today}

\begin{document}

\title{Protostars in the Elephant Trunk Nebula}
%\subtitle{}

\author{William T. Reach\altaffilmark{1}
Jeonghee Rho\altaffilmark{1}
Erick Young\altaffilmark{3}
James Muzerolle\altaffilmark{3}
Sergio Fajardo-Acosta \altaffilmark{1}
Lee Hartmann\altaffilmark{4}
Aurora Sicilia-Aguilar\altaffilmark{4}
Lori Allen\altaffilmark{4}
Sean Carey\altaffilmark{1}
Jean-Charles Cuillandre\altaffilmark{6}
Thomas H. Jarrett\altaffilmark{2}
Patrick Lowrance\altaffilmark{1}
Anthony Marston\altaffilmark{5}
Alberto Noriega-Crespo\altaffilmark{1}
Robert L. Hurt\altaffilmark{1}
}

\altaffiltext{1}{{\it Spitzer} Science Center, MS 220-6, 
California Institute of Technology,
Pasadena, CA 91125}

\altaffiltext{2}{Infrared Processing and Analysis Center, MS 100-22, 
California Institute of Technology,
Pasadena, CA 91125}

\altaffiltext{3}{Steward Observatory, University of Arizona, Tucson AZ}

\altaffiltext{4}{Smithsonian Astrophysical Observatory,
60 Garden St., Cambridge, MA 02138}

\altaffiltext{5}{European Space Research and Technology Center, Noordwijk, Netherlands}

\altaffiltext{6}{Canada-France-Hawaii Telescope, HI}

\email{reach@ipac.caltech.edu}

\begin{abstract}
The optically-dark globule IC 1396A is revealed using {\it Spitzer}
images at 3.6, 4.5, 5.8, 8, and 24 $\mu$m to be infrared-bright 
and to contain a set of previously unknown protostars. The
mid-infrared colors of the 24 $\mu$m detected sources indicate
several very young (Class I or 0) protostars and a dozen 
Class II stars. Three of the new sources (IC 1396A:$\gamma$, $\delta$,
and $\epsilon$) emit over 90\% of their bolometric luminosities
at wavelengths greater than 3 $\mu$m, and they are located within
$\sim 0.02$ pc of the ionization front at the edge of the globule.
Many of the sources have spectra that are still rising at 24 $\mu$m.
The two previously-known young stars LkH$\alpha$ 349
a and c are both detected, with component c harboring a massive
disk and component a being bare. Of order 5\% of the mass of material
in the globule is presently in the form of protostars in the
$10^5$--$10^6$ yr age range. This high star formation rate was
likely triggered by radiation from a nearby O star.
\keywords{stars: formation, ISM: globules, ISM: Individual IC 1396A,
infrared: stars, stars: individual LkH$\alpha$ 349
}

\end{abstract}
                            
\section{Introduction}

To demonstrate the capabilities of the new {\it Spitzer}
Space Telescope \citep{werner}, the facility scientists 
observed some targets that are
dark at optical wavelengths but expected to be bright in the infrared.
The Elephant Trunk Nebula is a textbook example of a dark globule 
immersed in an \ion{H}{2} region: see Figs. 7.3 and
7.4 of \citet{osterbrock}. 
The \ion{H}{2} region is IC 1396, a large (100$^\prime$ diameter) ionized
region within an infrared-bright shell. 
An excellent summary of knowledge of this region was written
by \citet{weikard}.
The distance to IC 1396 is estimated at 750 pc. 
Comparing the radio and H$\alpha$ brightness of the \ion{H}{2} region,
it appears there is only about 1.2 mag of spatially uniform 
foreground extinction \citep{morbidelli}. 
The region is ionized by the O6 star HD 206267, and it is joined by
20 B stars in the cluster Trumpler 37.
Radiation and winds from these stars sculpt the dense molecular gas 
into a variety of shapes with a predominant cometary 
morphology where the dense head points roughly back toward the O6 star.

The most prominent globules are located in the western and
northern portions of the \ion{H}{2} region. 
The rims of these globules are bright in H$\alpha$, as seen
in red optical images,
demarking the gradual progress of the ionization front from the O6 star 
into the globule.
The surfaces of the globules are also illuminated by 
scattered starlight, most easily visible in blue optical images.
A three-color optical image of the Elephant Trunk Nebula
from the Canada-France-Hawaii Telescope (CFHT)
is shown in the inset of Plate 1.
These globules were notable for being dark, but the combination
of dense material (required to harbor enough dust to extinguish all
background starlight) and a source of radiation and pressure (required
to make the bright rims) makes the globules very likely locations
of recent star formation.
The swept-back appearance of the globules is related to radiation and
wind pressure from the central O star. The globules are very bright
in CO emission \citep{patel} and
have atomic `tails' that extend up to 6 pc
opposite the central star \citep{moriarty}.
Two globules, IC 1396N and IC 1396W, exhibit recent
star formation with outflows and new Herbig-Haro objects 
\citep{reipurth,froebrich}.
The Elephant Trunk Nebula, IC 1396A, contains two T Tauri stars 
\citep{cohenkuhi,herbig}.
The new observations presented here reveal prolific, recent star formation
by uncovering a host of new, infrared-bright protostars.

\begin{figure}[th]
\plotone{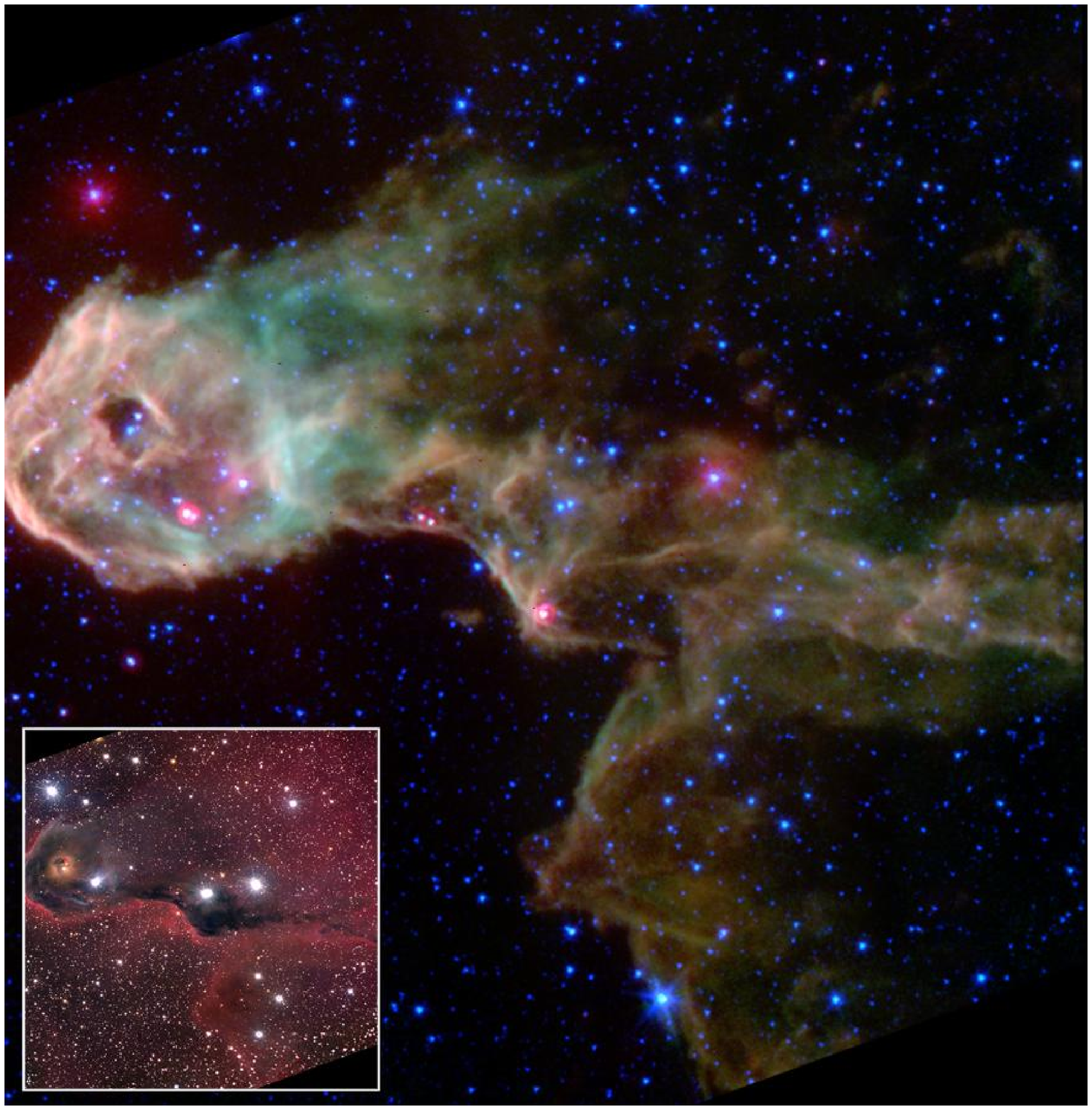}
\epsscale{1}
\figcaption[f1.eps]{[COLOR PLATE] 
{\it Spitzer} image of the Elephant Trunk Nebula in IC 1396,
combining the MIPS 24 $\mu$m (red) and IRAC 5.8+8 $\mu$m (green) and 
3.6+4.5 (blue) $\mu$m images. The inset shows the optical
image, combining red, green, and blue images from the CFHT.
This figure is available as an animation in the electronic 
edition of the {\it Astrophysical Journal}. The animation 
begins with the CFHT image and transforms into the {\it Spitzer}
image with each time-step moving to progressively longer wavelength.
North is rotated $13^\circ$ from vertical. The total horizontal
image size is $17.6'$.
\label{plate}}
\end{figure}
%\clearpage

\begin{figure}[th]
\epsscale{1}
\plotone{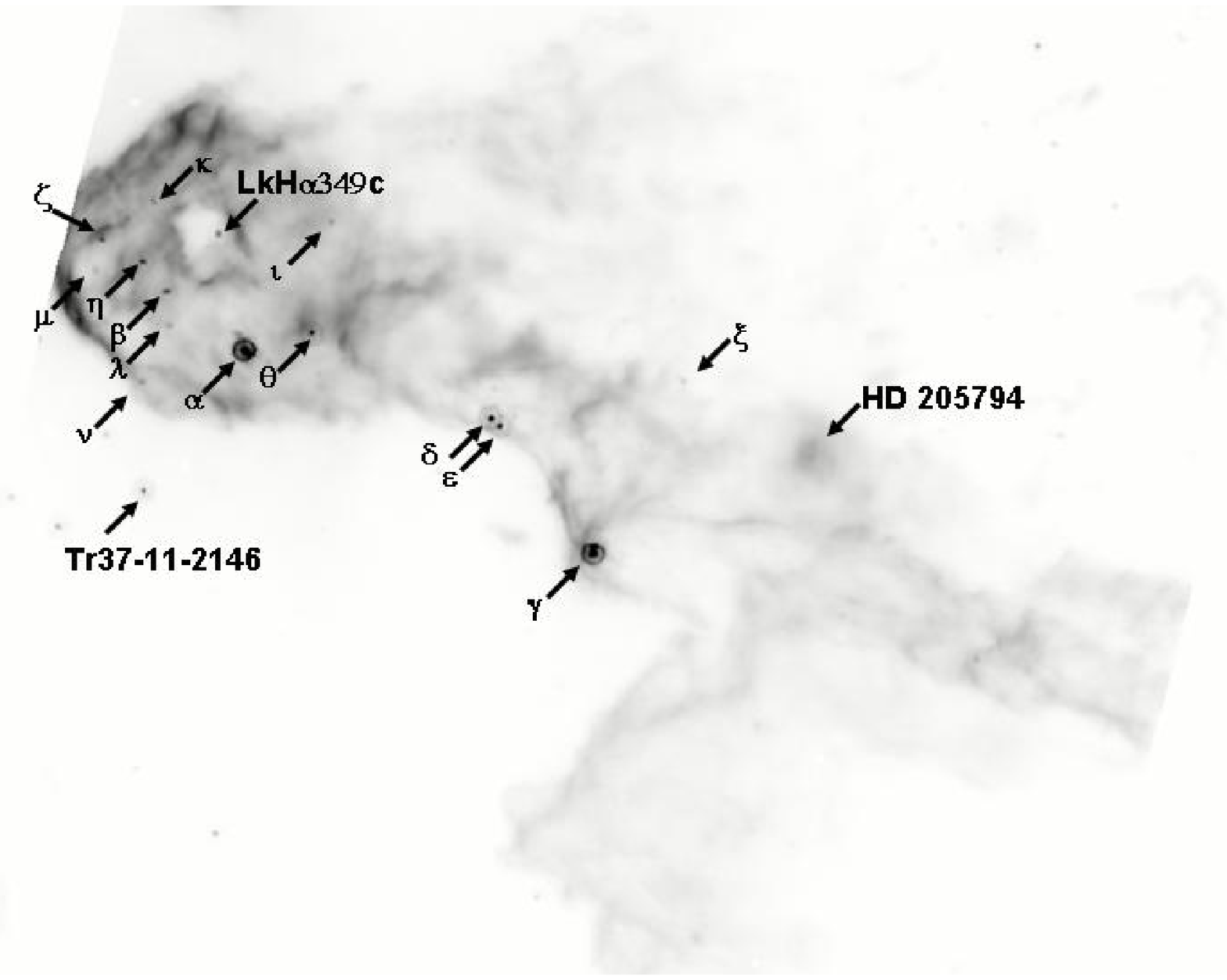}
\figcaption[f2.eps]{{\it Spitzer}/MIPS 24 $\mu$m image of 
the Elephant Trunk Nebula with 
newly discovered sources ($\alpha$--$\xi$) labeled.
North is up and east is left.
\label{lab}}
\end{figure}

\section{Observations}

IC 1396 was observed at 3.6--8 $\mu$m
on 2003 Nov 5 using IRAC \citep{fazio} [ads/sa.spitzer\#0006051840].
The IRAC observations have 2 sec frames and the map grid has 1/2-array
spacings in each direction, yielding a total observing time per sky position
of 8 sec; the IRAC map took 21 min of wall-clock time.
IC 1396 was observed at 24 $\mu$m on 2003 Nov 24 using MIPS \citep{rieke}
[ads/sa.spitzer\#0006052096].
The MIPS observations have 3 sec frames and used the small-field photometry
mode with a $3\times 3$ raster, 
yielding a total observing time per
sky position of 42 sec; the MIPS map took 17 minutes.
The photometric accuracy of the IRAC and MIPS data is better than 10\%.
For the point source photometry, {\it Spitzer} and 2MASS magnitudes are 
accurate to better than 0.1 and 0.05 magnitudes, respectively.

\section{Dust and gas in the globule}

The mid-infrared images reveal the body of the globule,
which is dark in optical photographs, as brilliant, structured emission
(Plate 1).
The inner edge of the rim of the globule, dark in the optical 
images, is bright in the infrared. The bright eastern
rim faces the O6 star.
%The extended emission has a morphology often described as `cometary'
%with brighter and more compact structure on the east with fainter 
%connected emission tailing off to the west. 
Within the head of the bright globule, there is a circular cavity
containing LkH$\alpha$ 349; 
that this is a cavity is known from 
microwave CO spectroscopy \citep{nakano}.
%In addition to the prominent
%globule that dominates the image, there are numerous much smaller ones,
%especially in the northeastern portion of the image.

The brightness, subtracting a background in a faint part of
the image, ranges from 4.5--31 MJy~sr$^{-1}$ at 24 $\mu$m,
from individual filaments in the south to the globule rim in the east. 
The ratio of the IRAC 8 $\mu$m to MIPS 24 $\mu$m surface
brightness ($I_\nu$) ranges from 1.4 to 2.6,
with the lower ratio being common and the higher ratio applying in
the region just west of the densest part of the globule (green
region in Fig.~\ref{plate}). For comparison, the diffuse
interstellar spectrum, convolved with {\it Spitzer} wavebands,
has a ratio $\sim 1.2$ \citep{ingalls},
similar to the bulk of the globule
emission (brownish red in Fig.~\ref{plate}). 
The other color ratios are $I_\nu(3.6, 4.5, 5.8)/I_\nu(8)=0.1$, 0.05, 
0.3.

The mid-infrared emission is a combination of small grains, polycyclinc
aromatic hydrocarbon (PAH), and H$_2$ lines.
Archival ISOCAM imaging spectroscopy from 5--17 $\mu$m 
shows that the IRAC 5.8 and 8 $\mu$m bands are dominated by 
PAH features at 6.2 and 7.7 $\mu$m typical for interstellar 
lines of sight \citep{boulanger}.
The ISOCAM spectra also detect pure rotational H$_2$ lines,
which contribute $\sim 2$\% 
and $\sim 8$\% in the 8 $\mu$m and 5.8 $\mu$m bands, respectively.
At shorter wavelengths, using a simple model 
based on IC~443 molecular shocks \citep{rho443},
we estimate H$_2$ lines contribute $\sim 50$\% and $\sim 30$\% 
of the 4.5 $\mu$m and 3.6 $\mu$m bands, respectively.
Though the H$_2$ excitation is uncertain, it appears
the 4.5 $\mu$m band could be a good tracer of H$_2$ emission.
That the 4.5 $\mu$m
is similar to the others suggests the H$_2$ and dust emission
are reasonably well correlated. However, some of the mid-infrared 
color variation in Figure~\ref{plate} is 
likely due to the relative brightness of H$_2$ compared to dust emission.
%The green regions in Figure~\ref{plate} could in principle
%have enhanced PAH or H$_2$ emission. 

\def\extra{
Ultraviolet observations showed that dust associated with the 
Tr 37 cluster is unusual, with a steep far-UV rise more typical 
of lines of sight to the Small Magellanic Cloud (SMC), perhaps due to
shattering of large grains by multiple supernovae \citep{clayton}.
However, the molecular material that we are observing seems not to be 
anomalous.
The situation is similar in the SMC, where molecular
clouds shows a Milky Way-like extinction curve \citep{lequeux} and
infrared spectrum \citep{reachsmc}.
}

\section{Stars in and near the globule}

The optically-bright star inside the cavity in IC 1396A
is LkH$\alpha$ 349a. This star may have cleared the central
cavity in its outflow phase, and it now excites a small reflection 
nebula.
One study found this star is of intermediate mass ($>3 M_\odot$), 
with a luminosity of 84 $L_\odot$ \citep{hessman}.
Optical and near-infrared photometry are consistent with pure
photospheric emission from this star. The new Spitzer data
extend this photospheric emission out to 8 $\mu$m and do
not detect the star at all at 24 $\mu$m: we conclude the star
has no circumstellar disk. 

A second pre-main-sequence star resides within the cavity with
LkH$\alpha$ 349a; we will refer to it as LkH$\alpha$ 349c,
and it was first noted by \citet{cohenkuhi}. While LkH$\alpha$ 349c
is 1.5 mag fainter than LkH$\alpha$ 349a in the visible, it is much 
brighter in the infrared. A small excess over photosphere is evident 
at 3.6 $\mu$m; it is very significant (1.9 mag) at 8 $\mu$m; and by 24 $\mu$m
there is negligible emission from the photosphere. 
Thus these two pre-main-sequence
stars are in very different stages of evolution, with LkH$\alpha$ 349a
having cleared its disk and LkH$\alpha$ 349c still harboring a
massive disk. Comparing optical photometry to \citet{siess}
%Siess, Dufour and Forestini
tracks yields an age of 2.0 Myr for LkH$\alpha$ 349c.
The star Tr 37 11-2146, labeled in Figure~\ref{lab}, is also detected
with a strong infrared excess suggesgting its youth. Ultraviolet observations
also reveal a strong excess, and the inferred accretion rate 
$\sim 10^{-7} M_\odot$~yr$^{-1}$ \citep{aurora}.
Most low-mass Tr 37 cluster members are 3--5 Myr old \citep{aurora};
if infrared excess is indicative of youth, then 
we expect the infrared-selected sample to be younger.

Three B stars are located in or near the globule: 
HD 205948 (B2V) is just north of the globule head, and 
HD 205794 (B0.5V) is west of the globule head, and HD 239710 (B3V)
is projected in front of the globule, just east of IC 1396:$\theta$.
All of these stars are surrounded by 24 $\mu$m nebulosity,
with in-band luminsoity $\sim 0.04 L_\odot$,
though the stars themselves are not detected at 24 $\mu$m.
We interpret these nebulae as molecular cloud dust heated
by the stars. From \citet{spitzer}, the the dust is
expected to be heated mostly by direct stellar radiation 
(and only a small contribution from scattered Ly$\alpha$), with 
Str\"omgren sphere radii $\sim 0.01$--$0.1 (10^3/n)^{2/3}$ pc, 
comparable to the observed nebula sizes.
That the \citet{spitzer}
models roughly correspond to the observations suggests {\it Spitzer}
is a very sensitive instrument for identifying B-star \ion{H}{2}
regions/reflection nebulae.

\section{Protostars}

\begin{figure}[th]
\epsscale{1}
\plotone{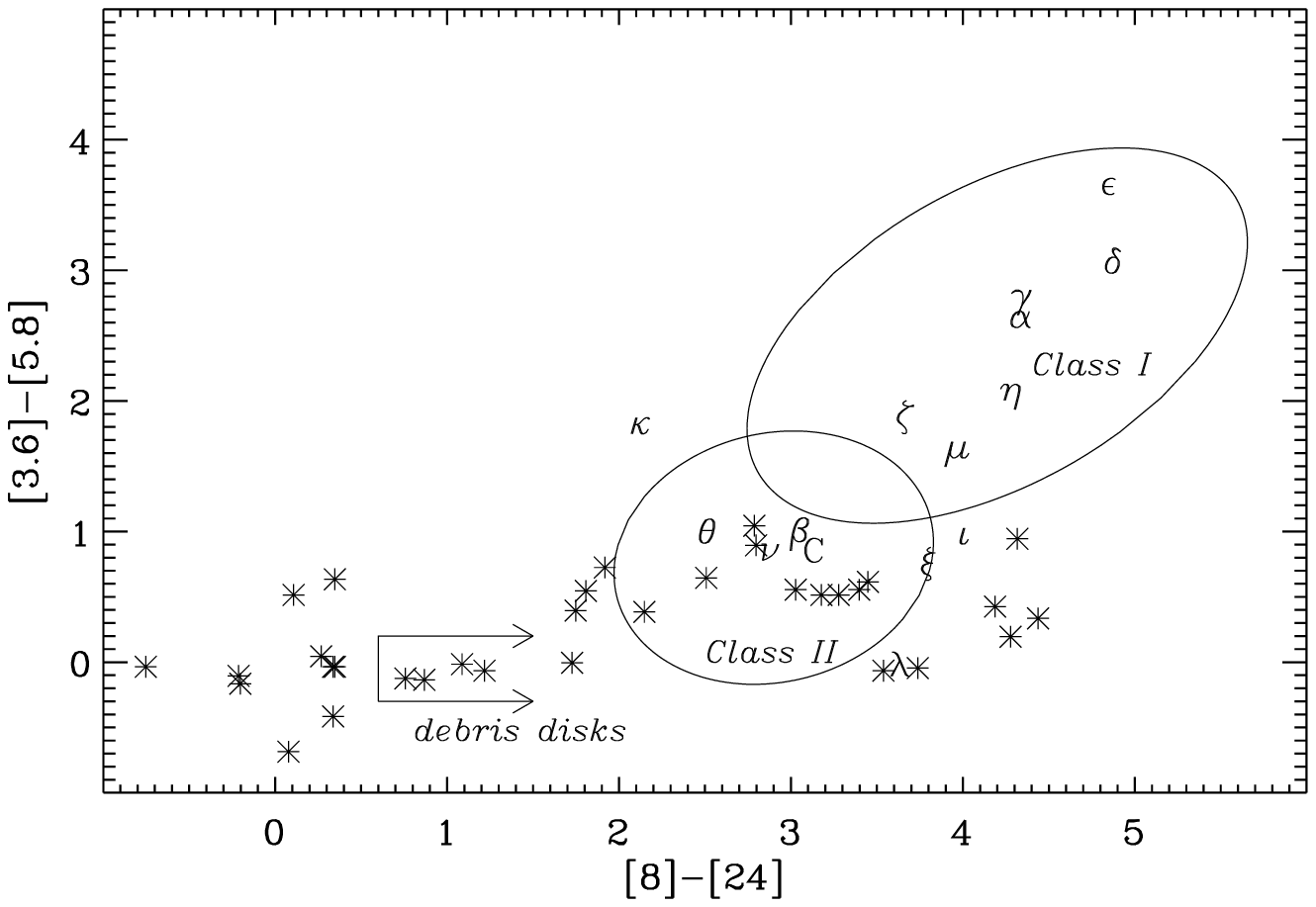}
\figcaption[f3.ps]{
{\it Spitzer} color-color diagram for 24 $\mu$m-detected sources in
the Elephant
Trunk Nebula (IC 1396A). Labeled points are from Table~\ref{phottab}.
The approximate locations of Classes I and II protostars and main
sequence stars with debris disks are outlined.
\label{colcol}}
\end{figure}

One of the most remarkable aspects of the {\it Spitzer} images is the presence
of numerous, bright mid-infrared sources with faint or no counterpart
at shorter wavelengths.
To find near-infrared counterparts, we used post-survey data from the 
2MASS project, which observed the
Elephant Trunk Nebula to a depth 6-times the nominal survey.
%\citep{tmass6xref}.
The brightest optical and 2MASS stars are seen in all the IRAC images, and 
even relatively faint stars at 3.6 $\mu$m. 
But the correspondence between optical and near-infrared sources and
8 $\mu$m sources is poorer, and by 24 $\mu$m the field is unrecognizable.
% but cross-referencing 
%to 2MASS (when there were counterparts) was straightforward using 
%IRAC 8 $\mu$m as the primary source list.
%{\it Spitzer} has revealed a new population of sources. 
There are four {\it IRAS} sources in the region, all `detected' at 25 $\mu$m.
One of these corresponds to the brightest {\it Spitzer} source $\alpha$,
but with 25 $\mu$m flux twice as high (due to globule contamination). 
Another is the infrared-bright 
T Tauri star that is the southernmost bright source in the
{\it Spitzer} image. The other two {\it IRAS}
sources are spurious; the {\it Spitzer} image shows bright but diffuse
24 $\mu$m emission.
%All of the sources in Table~\ref{phottab} are likely
%to be much brighter (1-2 orders of magnitude) than the {it IRAS} sensitivity 
%limit in the far-infrared, but they could not be discerned from the globule
%at low resolution. 

Figure~\ref{colcol} shows the {\it Spitzer} color-color diagram
for the 24 $\mu$m-detected sources. 
Few of the sources are dominated by stellar photospheres, which would
concentrate near [8]-[24]=[3.6-5.8]=0 since the foreground extinction
is small.
The sources to the upper-right are the reddest and
likely youngest protostars. 
There is a sequence of these protostars
progressing roughly toward the main sequence but noticeably offset
such that upon arriving at photospheric [3.6]-[5.8] color, the
young stars still retain significant disks.
A separate set
of sources has photospheric [3.6]-[5.8]$\sim 0$
but significant 24 $\mu$m excess; these are 
`debris disk' candidates. Some of these may actually be 
young stars with massive disks with inner holes---thus lacking 
the warmer dust that produces the shorter-wavelength emission.
Others are main sequence stars with debris disks due to collisional
comminution of their asteroids and comets.
For reference, the debris disk around $\beta$ Pic has 
[3.6]-[5.8]=0.0, [8]-[24]=3.3.
Some of the sources with extremely large 24 $\mu$m excesses, are 
actually contamination by globule material;
inspecting some of the points in the lower-right portion of the 
indicates their 24 $\mu$m emission is likely globule contamination.

\begin{figure}[th]
\epsscale{1}
\plotone{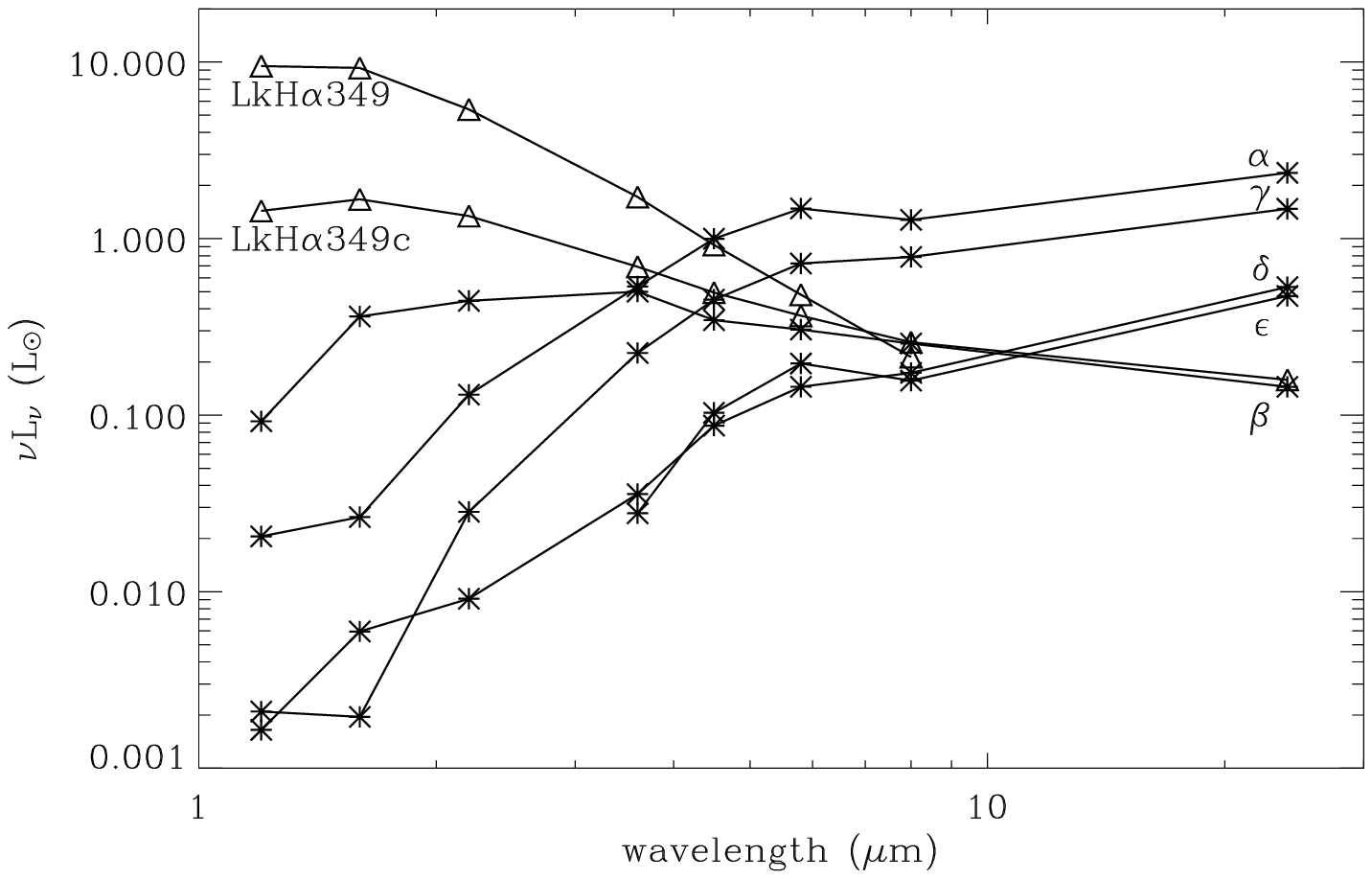}
\figcaption[f4.eps]{Spectral energy distributions for 5 newly-discovered
protostars ($\alpha$--$\epsilon$) and the two previously-known
young stars (LkH$\alpha$ 349 a and c). The spectrum of LkH$\alpha$ 349a
provides a reference for photospheric emission, while the others
show varying degrees of infrared excess.
\label{sedplot}}
\end{figure}

Fourteen potential protostars were studied in detail; they
are labeled with Greek letters in Figure~\ref{lab} and listed 
in Table~\ref{phottab}. 
The sources were selected first by inspecting
the 24 $\mu$m image, then by verifying that none
of the automatically-detected sources were missed.
IRAC magnitudes were calculated using a 5.4$''$ radius aperture
with respect to a reference annulus from 6 to 9$''$. An aperture correction
of 1.07, 1.08, 1.08, and 1.10 was applied to 3.6, 4.5, 5.8, and 8 $\mu$m,
respectively, to raise values to the standard calibration aperture of 12$''$.
MIPS magnitudes are within an aperture of 12$''$ with an aperture 
correction of 1.15.
Mid-infrared luminosities, $L_{MIR}$ ($L_\odot$), 
were calculated by integrating 
the spectral energy distribution from 3.6 to 24 $\mu$m (with linear 
interpolation between the observed wavelengths).
We measured the full width at half-maximum (FWHM) intensity of each source at
24 and 3.6 $\mu$m, to verify they are point-like at the intstrumental
resulution. All sources in Table~\ref{phottab} are
unresolved at 24 $\mu$m, with 5.2$''$ median FWHM equal to that
of the PSF.
The IRAC 3.6 $\mu$m point spread 
function has a FWHM of 1.6$''$  in a drizzled reconstruction at 0.3$''$
pixel scale.
All but two sources are between 1.5 and 2.1$''$, consistent with
being true point sources. The remaining two sources, $\alpha$
and $\xi$, are 2.4$''$ in
apparent size, but they are both blended with nearby stars and are
probably not truly resolved. 

Figure~\ref{sedplot} shows the spectral energy distributions of 
5 new protostars 
and the two T Tauri stars. The sources form a relatively clear sequence
with the mid-infrared emission decreasing as the near-infrared emission
increases, obviously suggestive of an evolutionary sequence wherein
the stars are gradually dispersing their envelopes. 
Comparing to the colors of pre-main sequence objects in the
Taurus-Auriga molecular cloud, we classified the objects as per
the criteria in \citet{kenyonhartmann} and \citet{wilking}.
Eight of the sources in Table~\ref{phottab} are Class 0, I 
or flat-spectrum protostars.
Eight of the sources in Table~\ref{phottab} (including
LkH$\alpha$ 349c and another star), and 47 overall, have mid-infrared
colors of Class II objects. 
The spectral energy distribution of source $\beta$ in 
Figure~\ref{sedplot} clearly shows a faint, barely-emerging photosphere.
The 8 $\mu$m data points may be slightly depressed due to a silicate absorption
feature, which can be very deep for protostars \citep{gibb}; 
spectroscopy is needed to understand these objects in more detail.
Nearly all of the {\it Spitzer}-detected objects 
are in very early stages, where the direct luminosity
from the photosphere is dwarfed by that of its envelope or disk.

\section{Discussion and Conclusions}

The new {\it Spitzer} images of the Elephant Trunk Nebula significantly 
alter our understanding of its nature from a dark, cold globule containing 
only two T Tauri stars into a cauldron of protostars. The mass of molecular
gas in the globule was estimated from independent
observations to be 
$220 M_\odot$ from a high-resolution CO map \citep{patel},
$170 M_\odot$ from the $^{12}$CO and 
$90 M_\odot$ from $^{13}$CO \citep{weikard},
and $< 30 M_\odot$ from a coarse $^{13}$CO map \citep{dobashi}; 
we will adopt $\sim 200 M_\odot$.
The globule mass is much less than the Virial mass,
estimated at 300--800 $M_\odot$ \citep{patel,weikard}, 
suggesting it would be quiescent and transient and {\it not} 
subject to gravitational collapse if it were in isolation---instead,
it would dissipate in a crossing time $\sim 5\times 10^5$ yr.
Including the pressure of the surrounding material only 
slightly increases the lifetime.

While we do not know the masses of the individual protostars,
we can get an order-of-magnitude estimate of their total mass
by assuming the average protostar is 0.5 $M_\odot$, yielding
a current protostellar mass of $\sim 8$--30 $M_\odot$, where the
lower estimate includes only Class I sources and the upper limit
includes Class II. 
The fraction of the globule currently in detected protostars is 4--15 \%,
similar to other emdebbed clusters \citep{ladalada}.
Assuming an age $\sim 10^5$--$10^6$ yr for the Class I--II protostars 
\citep{andrecore},
the star formation rate $\dot{M}_*\sim 10^{-4} M_\odot$~yr$^{-1}$,
with the relative populations of Class I and II consistent with their
ages and spatial distributions. (The Class I sources are mostly located
in the dense globule and in particular along its rim, while the
Class II sources are more widely distributed.)
The globule could supply material to form stars at
this rate for only $M/\dot{M}_*\sim 2\times 10^6$ yr,
comparable to the lifetime for photoevaporation of the globule
by the O6 star, $\sim 2\times 10^6$ yr \citep{LL94}. 
Coincidentally, the globule lifetimes, comparable to the 
main-sequence lifetime  of the O6 star.

The location of protostars---within a globule that is clearly being
illuminated and shaped by a nearby O star---and the high specific star
formation rate and concentration of so many young objects in 
a small volume all suggest that the Elephant Trunk Nebula protostars were
triggered to form in a cloud that may not have formed many
stars had it not been acted upon by external forces. 
The new protostars IC 1396A:$\gamma$ and IC 1396A:$\delta$
and IC 1396A:$\epsilon$ in particular have extremely red
mid-infrared spectral energy distributions indicating their youth.
The generation
of stars we are presently witnessing is likely to be the last produced
by this globule, because it will likely be photoevaporated by the
nearby O6 star before more stars can form.

\def\extra{
{\it Spitzer} is a fantastic new tool for studying the early stages
of star formation. A significant number of young protostars was first 
measured by the {\it Infrared Astronomical Satellite},
which was limited due to sensitivity and low angular resolution (which
confuses protostars with extended emission and other stars) to nearby
complexes such as Taurus \cite{irasref}. Using {\it Spitzer}, in particular
the IRAC 8 $\mu$m and MIPS 24 $\mu$m bands, we can detect protostars
to much greater distances and in a wide variety of environments.
}

\acknowledgements  

This work is based in part on observations made with the {\it Spitzer Space
Telescope}, which is operated by the Jet Propulsion Laboratory, California
Institute of Technology under NASA contract 1407. Support for this work was
provided by NASA through an award issued by JPL/Caltech. 
This publication makes use of data products from the Two Micron All Sky
Survey, which is a joint project of the University of Massachusetts and the
Infrared Processing and Analysis Center/California Institute of Technology,
funded by the National Aeronautics and Space Administration and the National
Science Foundation.

\clearpage

%\begin{table}
\begin{deluxetable}{lcccrrrrrrrrc}
\tablecaption{Photometry of young stellar objects in the Elephant Trunk Nebula\label{phottab}}
\tablewidth{7truein}
\tabletypesize{\footnotesize}
\tablehead{
\colhead{name}  & \colhead{Class} & \colhead{RA [21$^h$]}
& \colhead{Dec [57$^\circ$]}& \colhead{[24]} & \colhead{[8] }  
& \colhead{[5.8]} & \colhead{[4.5]} & \colhead{[3.6]} & 
\colhead{K$_s$} & \colhead{H}     & \colhead{J}  & \colhead{$L_{MIR}$}
}
\startdata
$\alpha$  & I & 36$^m$46.5$^s$ & 29$'$38$''$ & 2.01 &  6.28 &  7.13 &  8.30 &  9.70 & 12.66 & 15.31 & 16.37 &  2.8\\
$\beta$   & II & 36$^m$55.1$^s$ & 30 29      & 5.04 &  8.03 &  8.85 &  9.46 &  9.77 & 11.33 & 12.47 & 14.74 &  0.5\\
$\gamma$  & I/0 & 36$^m$07.8$^s$ & 26 36       & 2.52 &  6.80 &  7.91 &  9.17 & 10.64 & 14.32 & ...   & ...   &  1.6\\
$\delta$  & I/0 & 36$^m$19.2$^s$ & 28 38       & 3.63 &  8.44 &  9.66 & 10.95 & 12.64 & 15.55 & ...   & ...   &  0.5\\
$\epsilon$& I/0 & 36$^m$18.2$^s$ & 28 31       & 3.75 &  8.55 &  9.33 & 10.77 & 12.91 &  ...  &  ...  &  ...  &  0.4\\
$\zeta$   & I & 37$^m$02.2$^s$ & 31 15       & 4.46 &  8.06 &  9.02 &  9.86 & 10.84 & 13.62 & 16.48 &  ...  &  0.5\\
$\eta $   & I & 36$^m$57.7$^s$ & 30 55       & 4.93 &  9.15 & 10.25 & 11.33 & 12.25 &  ...  &  ...  &  ...  &  0.2\\
$\theta$  & II & 36$^m$39.0$^s$ & 29 52      & 4.07 &  6.53 &  7.25 &  7.67 &  8.16 &  9.50 & 10.21 & 12.02 &  2.0\\
$\iota$   & II & 36$^m$36.8$^s$ & 31 32      & 5.63 &  9.60 & 10.19 & 10.74 & 11.10 & 12.01 & 12.58 & 13.82 &  0.2\\
$\kappa$  & II & 36$^m$56.4$^s$ & 30 00      & 8.10 & 10.17 & 10.76 & 11.71 & 12.51 & 13.46 & 15.00 & 17.67 &  0.1\\
$\lambda$ & If & 36$^m$54.8$^s$ & 30 00      & 5.91 &  9.48 & 10.20 &  9.92 & 10.09 & 13.77 & 15.72 &  ...  &  0.2\\
$\mu$     & I & 37$^m$02.9$^s$ & 31 51       & 7.47 & 11.36 & 12.60 & 12.29 & 14.17 &  ...  &  ...  &  ...  &  0.1\\
$\nu$     & I/II & 36$^m$57.8$^s$ & 29 10    & 5.77 &  8.58 &  9.43 &  9.92 & 10.21 & 15.83 & 14.17 & 13.41 &  0.3\\
$\xi$     & I/II & 35$^m$57.8$^s$ & 29 11    & 6.67 & 10.42 & 12.38 & 12.58 & 13.07 & 14.81 & ...   & ...   &  0.1 \\
Tr 37 11-2146 
%$*$\tablenotemark{a}
&II&36$^m$57.5$^s$&27 33 & 5.21 &  8.34 &  9.12 &  9.41 &  9.87 & 10.82 & 11.52 & 12.54 &  0.4\\
Lk H$\alpha$ 349a
%A\tablenotemark{b}
&III&36$^m$49.3$^s$ &31 21 & ...  &  8.22 &  8.35 &  8.39 &  8.43 &  8.62 &  8.95 &  9.71 &  0.7\\
Lk H$\alpha$ 349a
%C\tablenotemark{c}
&II &36$^m$50.6$^s$ &31 10 & 4.94 &  8.01 &  8.65 &  9.07 &  9.42 & 10.13 & 10.81 & 11.76 &  0.6\\
\enddata
%\tablenotetext{a}{*=Tr 37 11-2146 \citep{aurora}}
%\tablenotetext{b}{A=Lk H$\alpha$ 349a}
%\tablenotetext{c}{C=Lk H$\alpha$ 349c}
\end{deluxetable}

\end{document}